\documentclass[letter]{jpsj3}
\usepackage{txfonts}

\title{Fine-Tuning of Magnetic Interactions in Organic Spin Ladders}

\author{Hironori \textsc{Yamaguchi}$^1$\thanks{E-mail:yamaguchi@p.s.osakafu-u.ac.jp}, Hirotsugu \textsc{Miyagai}$^1$, Tokuro \textsc{Shimokawa}$^2$, Kenji \textsc{Iwase}$^1$, Toshio \textsc{Ono}$^1$, Yohei \textsc{Kono}$^3$, Naoki \textsc{Kase}$^3$\thanks{Present address: Faculty of Engineering, Niigata University, Niigata 950-2181, Japan}, Koji \textsc{Araki}$^3$, Shunichiro \textsc{Kittaka}$^3$, Toshiro \textsc{Sakakibara}$^3$, Takashi \textsc{Kawakami}$^4$, Kouichi \textsc{Okunishi}$^5$, and Yuko \textsc{Hosokoshi}$^1$\thanks{E-mail:yhoso@p.s.osakafu-u.ac.jp}}
\inst{$^1$Department of Physical Science, Osaka Prefecture University, Sakai, Osaka 599-8531, Japan \\
$^2$Department of Earth and Space Science, Faculty of Science, Osaka University, Toyonaka, Osaka 560-0043, Japan\\
$^3$Institute for Solid State Physics, the University of Tokyo, Kashiwa, Chiba 277-8581, Japan\\
$^4$Department of Chemistry, Osaka University, Toyonaka, Osaka 560-0043, Japan \\
$^5$Department of Physics, Niigata University, Niigata 950-2181, Japan} 

\abst{We have succeeded in synthesizing two types of new organic radical crystals 3-I-V [= 3-(3-iodophenyl)-1,5-diphenylverdazyl] and 3-Br-4-F-V [= 3-(3-bromo-4-fluorophenyl)-1,5-diphenylverdazyl].
Their crystal strucutures were found to be isomorphous to that of previously reported spin ladder 3-Cl-4-F-V.
Through the quantitative analysis of their molecular arrangements and magnetic properties, we confirmed that these materials form ferromagnetic chain-based spin ladders with slightly modulated magnetic interactions.
These results present the first quantitative demonstration of the fine-tuning of magnetic interactions in the molecular- based materials.
}

\begin{document}
\maketitle
Stable open-shell organic molecules have recently attracted considerable attention as next-generation spin sources for organic magnets~\cite{pNPNN, BIMNN}, radical-based batteries~\cite{battery1,battery2}, and spintronic devices~\cite{spintro1,spintro3}.
They are expected to facilitate the design of magnetic materials because of their wide range of chemical modifications and structural flexibility. 
However, since their molecular arrangements are quite sensitive to chemical modifications, a slight difference of a molecule often changes intermolecular magnetic interactions drastically, which results in the formation of different magnetic lattices. 
It is a crucial issue in the design of magnetic materials to understand the essential molecular structures that stabilize molecular arrangements.

A large number of organic radical materials have been synthesized in the past two decades through the use of representative free radicals such as nitroxide and nitronyl nitroxide. 
They have attracted considerable attention as ideal $S$ = 1/2 quantum spin systems owing to the low dimensionality of molecular-based material. 
Some of them have been reported to possess isomorphous structures with different chemical modifications, which indicates a potential to fine-tune intermolecular magnetic interactions~\cite{INN}. 
Recently, we established synthetic techniques for preparing high-quality verdazyl radical crystals and successfully synthesized a variety of materials forming unconventional magnetic lattices~\cite{3Cl4FV,2Cl6FV,26Cl2V, pBrV}.
The verdazyl radical can exhibit a delocalized $\pi$-electron spin density even on non-planar molecular structures, which is an essential property for forming unconventioanl magnetic lattices.
This flexibility of the molecular structure enabled us to design molecular arrangements by chemical modification, resulting in the fine-tuning of intermolecular magnetic interactions.

The spin ladder is one of the most representative quantum spin systems.
The $S$ = 1/2 antiferromagnetic (AFM) spin ladder, which exhibits AFM rung and leg interactions, has been extensively studied in relation to field-induced quantum phase transitions and high-$T_c$ superconductors~\cite{SL1,SLSC}. 
Recent investigations on new copper complexes, BPCB and DIMPY, have furthered the understanding of $S$ = 1/2 AFM spin ladders and their field-induced quantum liquid phases~\cite{BP_CM,DY_CM,DY_M}.
Furthermore, much attention has been given to the interesting contrast for the associated Tomonaga-Luttinger liquids between strong-leg and strong-rung coupling regimes, where attractive and repulsive interactions act between spinless fermions, respectively~\cite{BP_NMR,DY_NMR}.
In contrast to such AFM spin ladders, in ferromagnetic (FM) chain-based spin ladders, where FM chains are coupled by AFM rung interactions, it is uncertain whether a quantum state can be stabilized because of suppression of quantum fluctuations originating from the FM chains~\cite{1DFMth_1,1DFMth_2}.
We previously reported the first experimental realization of an $S$ = 1/2 FM chain-based spin ladder in 3-Cl-4-F-V~\cite{3Cl4FV}, where the ratio between the rung and the leg interactions, $\gamma=|J_{\rm{rung}}/J_{\rm{leg}}|$, was evaluated as 0.55. 
Because 3-Cl-4-F-V has strong-leg coupling, a model substance that has a strong-rung coupling with $\gamma \textgreater 0.55$ is required to examine the differences in coupling regimes.
Additionaly, further strong-leg couping with a small value of $\gamma {\textless} 0.55$ is also necessary to clarify the classical magnetic effect of the FM chains.

In this study, we present two types of new model materials for $S$ = 1/2 FM chain-based spin-ladder, 3-I-V [= 3-(3-iodophenyl)-1,5-diphenylverdazyl] and 3-Br-4-F-V [= 3-(3-bromo-4-fluorophenyl)-1,5-diphenylverdazyl], where we successfully modulated $\gamma$ by chemical modification of the 3-Cl-4-F-V molecule. 
The $ab$ $initio$ molecular orbital (MO) calculation indicated the formation of spin-ladders with slightly different magnetic interactions.
Indeed, we obtained $\gamma$ = 0.50 and 1.5 from quantitative analysises of the magnetic properties for 3-I-V and 3-Br-4-F-V, respectively.
These results provide the first quantitative demonstration of fine-tuning of magnetic interaction in organic radical materials and desirable model substance of $S$ = 1/2 FM chain-based spin ladder with different coupling regimes.

We synthesized 3-I-V and 3-Br-4-F-V using a conventional procedure similar to that used for preparing the typical
verdazyl radical 1,3,5-triphenylverdazyl~\cite{procedure}.
The crystal structure was determined by the intensity data collected using a Rigaku CCD Mercury diffractometer. 
The magnetic susceptibility and magnetization curves were measured using a commercial SQUID magnetometer (MPMS-XL, Quantum Design) and a capacitive Faraday magnetometer, respectively.
The experimental results were corrected for the diamagnetic contributions -1.86${\times}$ $10^{-4}$ and -2.56 ${\times}$ $10^{-4}$ emu mol$^{-1}$, which are determined based on the QMC analysis (to be described) and close to those calculated by Pascal's method, for 3-I-V and 3-Br-4-F-V, respectively.
The specific heat was measured using a hand-made apparatus by a standard adiabatic heat-pulse method for 0.30 K $<T<$ 2.5 K. 
Considering the isotropic nature of organic radical systems, all experiments were performed using randomly oriented small single crystals.

Figures 1(a) and 1(b) show the molecular structures of 3-I-V and 3-Br-4-F-V, respectively.
For 3-I-V, the crystallographic parameters are as follows: monoclinic, space group $P2_{1}/n$, $a$ =  4.905(2) $\rm{\AA}$, $b$ = 33.149(13) $\rm{\AA}$, $c$ = 10.964(4) $\rm{\AA}$, $\beta$ = 94.091(5)$^{\circ}$~\cite{3IV}, whereas for 3-Br-4-F-V, they are as follows: monoclinic, space group $P2_{1}/n$, $a$ =  5.1529(16) $\rm{\AA}$, $b$ = 32.894(10) $\rm{\AA}$, $c$ = 10.343(3) $\rm{\AA}$, $\beta$ = 96.266(4)$^{\circ}$~\cite{3Br4FV}.
These results indicate that the molecular arrangements of these materials are isomorphous to that of 3-Cl-4-F-V~\cite{3Cl4FV}.
The verdazyl ring with four nitrogen atoms, the right side two phenyl rings, and the left side phenyl ring are labeled ${\rm{R}_{1}}$, ${\rm{R}_{2}}$, ${\rm{R}_{3}}$, and ${\rm{R}_{4}}$, respectively, for both molecules, as shown in Figs. 1(a) and 1(b).
It is significant that both molecules have non-planar structures, and their dihedral angles are listed in Table I with those of 3-Cl-4-F-V.
In these materials, the large covalent radius of the halogens introduced at the 3-position, which correspond to I, Br, and Cl atoms in 3-I-V, 3-Br-4-F-V, and 3-Cl-4-F-V, respectively, result in relatively large dihedral angles only in ${\rm{R}_{1}}$-${\rm{R}_{3}}$, while the other dihedral angles are comparatively small. 
These results crucially show that the halogens introduced at the 3-position define the dihedral angles in the molecules.

Here, attention is directed mainly toward the structural features related to the R$_1$$\sim$R$_3$ rings, which have relatively large spin densities~\cite{3Cl4FV,2Cl6FV,26Cl2V}.
Figure 1(c) shows the remarkable uniform chain structure along the $a$-axis in 3-I-V, where the C-C and N-C short contacts less than 3.6 $\rm{\AA}$ are labeled $\rm{d}_1$-$\rm{d}_3$ and listed in Table I.
The corresponding contacts for 3-Br-4-F-V and  3-I-V are slightly longer and shorter than those in 3-Cl-4-F-V, respectively. 
In addition, these molecular arrangements along the $a$-axis have slight differences in their overlaps, as shown in Figs. 2(a)-2(c).
We can make the differences easily understood in comparison with the case of 3-Cl-4-F-V, where the electrostatic repulsion between the Cl and F atoms largely defines the relative configuration of the molecular pair, as shown in Fig. 2(b). 
For 3-Br-4-F-V, the Br atom with a larger covalent radius enhances the electrostatic repulsion, which results in the molecules moving slightly away from each other, as shown in Fig. 2(c). 
For 3-I-V, because there is no halogen at the 4-position, the two molecules move toward each other, as shown in Fig. 2(a).  
Figure 1(d) shows the molecular arrangement viewed along the uniform chain direction in 3-I-V.
The C-C short contacts labeled $\rm{d}_4$, which are doubled by an inversion symmetry, connect two neighboring chains to form a two-leg ladder structure, as shown in Fig. 1(e).
Because there are few differences in overlap of molecules between the three materials, the rung interactions are considered to be directly affected by these short distances. 
We performed $ab$ $initio$ molecular orbital (MO) calculations~\cite{MO} and found that the rung and leg interactions are dominant for both 3-I-V and 3-Br-4-F-V.
They are evaluated as $J_{\rm{rung}}/k_{\rm{B}}$ = 7.8 K and $J_{\rm{leg}}/k_{\rm{B}}$ = $-$9.4 K ($\gamma$ = 0.83) for 3-I-V, and $J_{\rm{rung}}/k_{\rm{B}}$ = 8.4 K and $J_{\rm{leg}}/k_{\rm{B}}$ = $-$4.8 K ($\gamma$ = 1.75) for 3-Br-4-F-V, which are defined in the Heisenberg spin Hamiltonian given by
\begin{equation}
\mathcal {H} = J_{\rm{leg}}{\sum^{}_{ij}}\textbf{{\textit S}}_{i,j}{\cdot}\textbf{{\textit S}}_{i+1,j}+J_{\rm{rung}}{\sum^{}_{i}}\textbf{{\textit S}}_{i,1}{\cdot}\textbf{{\textit S}}_{i,2},
\end{equation}
where $\textbf{{\textit S}}_{i,j}$ are the spin operators acting on site $i$ of leg $j$ = 1,2 of the ladder.
These results quantitatively demonstrate the formation of FM chain-based spin ladders with slightly modulated magnetic interactions through a perspective of molecular arrangement. 

To verify the actual magnetic state, we compare the magnetic properties of 3-I-V and 3-Br-4-F-V with those of 3-Cl-4-F-V.
Figure 3 shows the temperature dependence of the magnetic susceptibilities ($\chi$ = $M/H$). 
Above 100 K, the Curie-Weiss law is followed, $\chi$ = $C/(T-{\theta}_{\rm{W}})$.
The estimated Curie constants are about $C$ = 0.368 emu$\cdot$ K/mol, which is slightly smaller than the expected 0.375 emu$\cdot$ K/mol and indicates that the purity of the radicals is about 98 $\%$, for both 3-I-V and 3-Br-4-F-V.
We take these purities into consideration in all of the following calculations.
The Weiss temperatures are estimated to be ${\theta}_{\rm{W}}$ = +4.4(3) K and $-$0.8(2) K for 3-I-V and 3-Br-4-F-V, respectively.
The observed shift of the broad peak temperatures indicate the difference of the AFM $J_{\rm{rung}}$ values.
For 3-I-V, a broad peak of ${\chi}T$~\cite{3Cl4FV} caused by the relatively strong FM contribution appears at about 13 K, as shown in the upper inset of Fig. 3. 
For 3-Br-4-F-V, the small Weiss temperature indicates a weak mean-field due to competition between the FM and AFM interactions, as shown in the lower inset of Fig. 3.
Figure 4(a) shows the magnetization curves. 
The magnetization curve of 3-Br-4-F-V exhibits an excitation gap of about 4.5 T.
Conversely, the magnetization curve of 3-I-V indicates the disappearance of the zero-field energy gap owing to the interladder interactions as seen in that for 3-Cl-4-F-V. 
Figure 4(b) shows the temperature dependence of the specific heats at the zero-field with the previously reported result of 3-Cl-4-F-V~\cite{3Cl4FV}, which indicates a phase transition to an ordered state at about $T_\mathrm{N}$ = 1.1 K. 
For 3-I-V, we found a distinct $\lambda$-type anomaly at about $T_\mathrm{N}$ = 1.4 K, which clearly indicates a phase transition to an ordered state.
This remarkable phase transition behavior indicates that the contribution of interladder interactions are enhanced in 3-I-V.

To further examine the magnetic properties, we carried out a quantitative analysis assuming an $S$ = 1/2 FM chain-based spin ladder.
We calculated the magnetic susceptibility as a function of $\gamma=|J_{\rm{rung}}/J_{\rm{leg}}|$ by using the quantum Monte Carlo (QMC) method~\cite{QMC}.
We found that there was no distinct $\gamma$ dependence of $\chi$, and ${\chi}T$ values are very sensitive to changes in $\gamma$, as shown in the insets of Fig. 3. 
We then obtained good agreement between experiment and calculation by using the parameters in Table II, as shown in Fig. 3.
Comparing general difference between experimental and the $ab$ $initio$ values, the obtained parameters can be regarded as consistent with those evaluated from the MO calculation.
The ground states for the models of 3-I-V and 3-Br-4-F-V are the rung-singlets with the excitation gaps $\Delta \approx$ 0.5 K and 6.9 K to the lowest triplet states, respectively.
Because the expected value of the energy gap is quite small for 3-I-V, the inter-ladder couplings cause a disappearance of the energy gap and phase transition to an ordered state, which results in the deviation of magnetic susceptibility in the low-temperature region. 
We calculated the magnetization curves using the same parameters as in the above analysis~\cite{DMRG}.
We obtained good agreement between experiment and calculation, while there are some deviations in the vicinity of the zero and saturation fields for 3-I-V, as shown in Fig. 4(a). 
For 3-Br-4-F-V, considering the agreement in the value of the magnetic field corresponding to the energy gap, the interladder interactions are inconsequential for the intraladder ones.
For 3-I-V, since the experimental temperature of 0.5 K is sufficiently lower than the $T_{\rm{N}}$, the difference between the calculated gapped and the actual gapless behaviors is enhanced.

We confirmed that the value of $J_{\rm{rung}}$ is directly affected by the short contact distance d$_4$ between molecules. 
In the case of the leg direction, there are some differences in molecular overlaps, as shown in Figs. 2(a)-2(c). 
The slight change in molecular overlaps is known to have relatively strong influence on the value of magnetic interaction. 
Although it is difficult to relate these differences to the value of $J_{\rm{leg}}$, we can evaluate the relation between the molecular overlaps and the magnetic interaction from the actual changes in $J_{\rm{leg}}$.
In the present case, the slight shift of molecular overlap away from each other makes $J_{\rm{leg}}$ weak, as seen in 3-Br-4-F-V, while molecular overlap in the direction of approach induces only a small change as seen in 3-I-V.

In summary, we successfully synthesized new organic radical crystals of 3-I-V and 3-Br-4-F-V, where we have modified the magnetic interactions of spin-ladder 3-Cl-4-F-V by chemical modulations.
The $ab$ $initio$ MO calculation actually indicated the formation of $S$ = 1/2 two-leg spin ladder with FM leg interactions.
We analyzed the magnetic properties by using the QMC and DMRG calculations and explained them as the expected spin ladders.
Three types of verdzyl radical crystals, 3-I-V, 3-Cl-4-F-V, and 3-Br-4-F-V, are revealed to be optimal models of $S$ = 1/2 FM chain-based spin ladders with both strong-leg and strong-rung coupling regimes.
Further experimental studies on their field induced phases will reveal quantum states in this FM chain-based system and clarify the classical magnetic effect of the FM chains, which would be significant information for advancing our understanding of the quantum effect in magnetic materials.    
Our results present the first quantitative demonstration of fine-tuning of magnetic interaction in molecular-based organic radical compounds and will promote synthesis of magnetic materials using similar types of organic radicals as verdazyl.

\begin{figure}[t]
\begin{center}
\includegraphics[width=17pc]{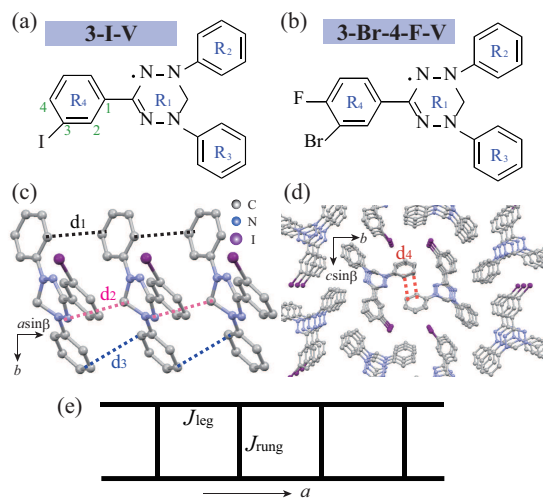}
\caption{(Color online) Molecular structures of 3-I-V (a) and 3-Br-4-F-V (b). Molecular arrangement of 3-I-V viewed along the $c$- (c) and $a$-axis (d). Broken lines indicate C-C and N-C short contacts. Hydrogen atoms are omitted for clarity. (e) Two-leg ladder along the $a$-axis formed by intermolecular interactions $J_{\rm{leg}}$ and $J_{\rm{rung}}$.}\label{f1}
\end{center}
\end{figure}

\begin{figure}[t]
\begin{center}
\includegraphics[width=21pc]{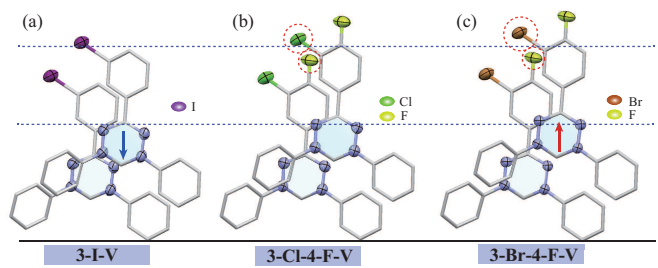}
\caption{(Color online) Neighboring molecules along the leg direction viewed in the direction perpendicular to the verdazyl rings for 3-I-V (a), 3-Cl-4-F-V (b), and 3-Br-4-F-V (c). Broken circles enclose halogens with relatively strong electrostatic repulsion determining these differences. Transverse lines bring the shift of the molecules into clear view. The arrows indicate the directions of shifts in comparison to the molecular overlap for 3-Cl-4-F-V.}\label{f2}
\end{center}
\end{figure}

\begin{figure}[t]
\begin{center}
\includegraphics[width=19pc]{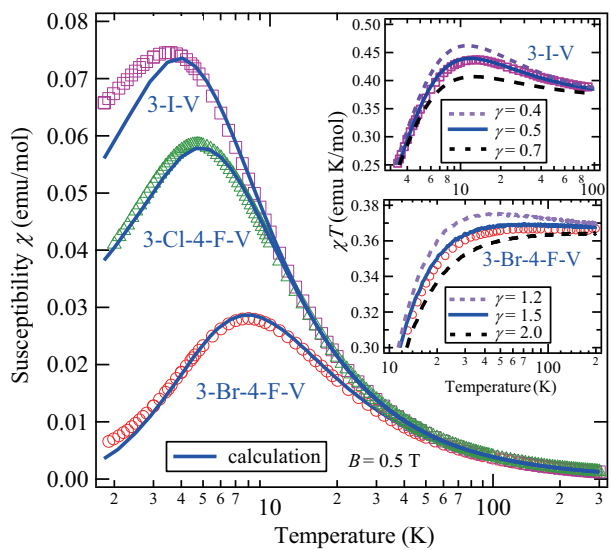}
\caption{(Color online) Temperature dependence of magnetic susceptibility $\chi$ ($M/H$) at 0.5 T for 3-I-V, 3-Cl-4-F-V~\cite{3Cl4FV}, and 3-Br-4-F-V. 
The upper and lower insets shows temperature dependence of ${\chi}T$ for 3-I-V and 3-Br-4-F-V, respectively. }\label{f2}
\end{center}
\end{figure}

\begin{figure}[t]
\begin{center}
\includegraphics[width=19pc]{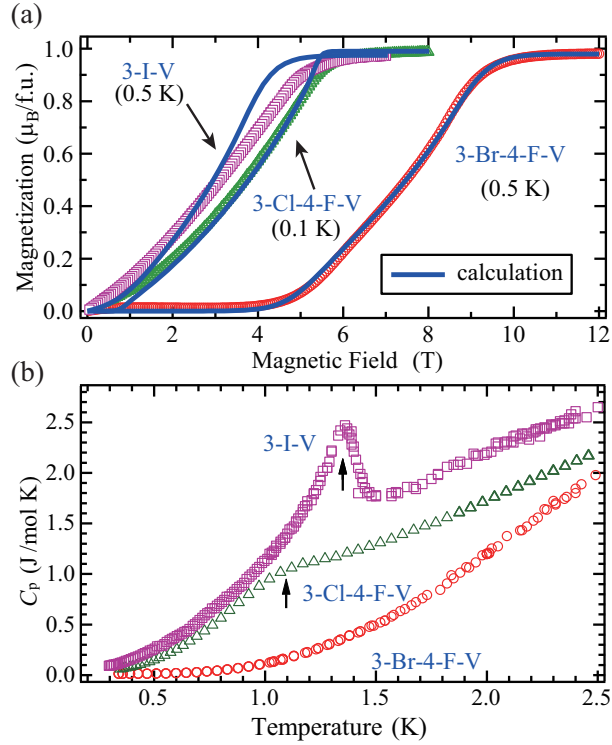}
\caption{(Color online) (a) Magnetization curves of 3-I-V and 3-Br-4-F-V at 0.5 K, and 3-Cl-4-F-V at 0.1 K~\cite{3Cl4FV}. 
The accompanying solid lines for 3-I-V, 3-Cl-4-F-V, and 3-Br-4-F-V represent the calculated results for $\gamma=|J_{\rm{rung}}/J_{\rm{leg}}|$ = 0.5, 0.55, and 1.5, respectively. 
(b) Temperature dependence of specific heat at the zero field for 3-I-V, 3-Cl-4-F-V~\cite{3Cl4FV}, and 3-Br-4-F-V. The arrows indicate phase transition temperatures.}\label{f2}
\end{center}
\end{figure}

\begin{table}
\caption{Dihedral angles and short intermolecular distances of 3-I-V, 3-Cl-4-F-V, and 3-Br-4-F-V.}
\label{t1}
\begin{center}
\begin{tabular}{cccccccc}
\hline 
material & ${\rm{R}_{1}}$-${\rm{R}_{2}}$ ($^{\circ}$)   &   ${\rm{R}_{1}}$-${\rm{R}_{3}}$ ($^{\circ}$)  &  ${\rm{R}_{1}}$-${\rm{R}_{4}}$ ($^{\circ}$)  & $\rm{d}_1$ ($\rm{\AA}$)   & $\rm{d}_2$ ($\rm{\AA}$)  & $\rm{d}_3$ ($\rm{\AA}$)  & $\rm{d}_4$ ($\rm{\AA}$)\\
\hline 
3-I-V & 16 & 37 & 16 & 3.54 & 3.47 & 3.43 & 3.75 \\
3-Cl-4-F-V  & 16& 34 & 17  & 3.58  & 3.54  & 3.50  & 3.69 \\
3-Br-4-F-V & 16 & 36 & 17 & 3.60 & 3.58 & 3.51 & 3.67 \\

\hline
\end{tabular}
\end{center}
\end{table}

\begin{table}
\caption{Evaluated magnetic parameters and observed phase transition temperatures of 3-I-V, 3-Cl-4-F-V, and 3-Br-4-F-V.}
\label{t1}
\begin{center}
\begin{tabular}{ccccccc}
\hline 
material   &   $|J_{\rm{rung}}/J_{\rm{leg}}|$   &  $J_{\rm{rung}}/k_{\rm{B}}$ (K) & $J_{\rm{leg}}/k_{\rm{B}}$ (K) & expected gap (K) & ${\theta}_{\rm{W}}$ (K) & $T_{\rm{N}}$ (K)\\
\hline 
3-I-V  & 0.50 & 5.8 & -11.6  & 0.5  & 4.4 & 1.4 \\
3-Cl-4-F-V & 0.55 & 7.3 & -13.3 & 1.1  & 4.5 & 1.1\\
3-Br-4-F-V & 1.5 & 12.5 & -8.3 & 6.9 & -0.8  & not observed \\

\hline
\end{tabular}
\end{center}
\end{table}

We thank T. Okubo, M. Yoshida, M. Takigawa, and T. Tonegawa for the valuable discussions. This research was partly supported by KAKENHI (Nos. 24740241, 24540347, and 24340075) and the Murata Science Foundation．A part of this work was performed under the interuniversity cooperative research program of the joint-research program of ISSP, the University of Tokyo. Some computations were performed using the facilities of the Supercomputer Center, the ISSP, The University of Tokyo.


\begin{thebibliography}{9}
\bibitem{MO}
The $ab$ $initio$ MO calculations were performed using the UB3LYP method as broken-symmetry hybrid density functional theory calculations.
All the calculations were performed using the Gaussian 09 program package, and the basis functions used were SDD and 6-31G for 3-I-V and 3-Br-4-F-V, respectively. 
To estimate the intermolecular magnetic interaction  of the molecular pairs within 4.0 $\rm{\AA}$, we applied our previously presented evaluation scheme~\cite{MOcal}.
\bibitem{QMC}
Our QMC code is based on the continuous-imaginary-time-loop algorithm~\cite{QMC1} and the directed loop algorithm in the stochastic series expansion representation~\cite{QMC2}. 
The calculation was performed for a system size of 96 under a periodic boundary condition using the ALPS application~\cite{ALPS}.
\bibitem{DMRG}
For the magnetization curve of 3-Br-4-F-V, finite-temperature DMRG calculations~\cite{DMRG1, DMRG2,DMRG3} were performed for the quantum transfer matrix under the periodic boundary condition in the imaginary time direction. 
The maximum number of the basis kept was 200, within which the computational results converged well.
\end{thebibliography}
\end{document}